\documentclass[sigconf,screen]{acmart}
\setcopyright{rightsretained}
\acmPrice{}
\acmDOI{10.1145/3540250.3560879}
\acmYear{2022}
\copyrightyear{2022}
\acmSubmissionID{fse22ivr-p9-p}
\acmISBN{978-1-4503-9413-0/22/11}
\acmConference[ESEC/FSE '22]{Proceedings of the 30th ACM Joint European Software Engineering Conference and Symposium on the Foundations of Software Engineering}{November 14--18, 2022}{Singapore, Singapore}
\acmBooktitle{Proceedings of the 30th ACM Joint European Software Engineering Conference and Symposium on the Foundations of Software Engineering (ESEC/FSE '22), November 14--18, 2022, Singapore, Singapore}

\newif\ifDEBUG
\DEBUGfalse

\newif\ifSPACEHACK
\SPACEHACKtrue
\usepackage{comment}
\usepackage{svg}
\usepackage{microtype}
\usepackage{comment}

\newcommand{\eg}{\textit{e.g.,\ }}
\newcommand{\etal}{\textit{et al.\ }}

\usepackage{mdframed}
\mdfsetup{skipabove=0.5\topskip,skipbelow=0.5\topskip,align=center}

\usepackage{amsthm}
\newtheorem{thm}{Theorem}%[section]
\usepackage{enumitem}
\setlist[itemize]{leftmargin=*,noitemsep,topsep=0pt}
\setlist[enumerate]{leftmargin=*}
 
\usepackage{todonotes}
\usepackage{soul}
 
\ifDEBUG
    \newcommand{\DA}[1]{\todo[color=white,inline]{DA:#1}}
    \newcommand{\JD}[1]{\todo[color=yellow,inline]{JD:#1}}
    \newcommand{\WJ}[1]{\todo[color=pink,inline]{WJ:#1}}
   
    \newcommand{\TODO}[1]{\hl{#1}}
\else
    \newcommand{\DA}[1]{}
    \newcommand{\JD}[1]{}
    \newcommand{\WJ}[1]{}
   
    \newcommand{\TODO}[1]{}
\fi
 
\usepackage{booktabs} % For formal tables
\usepackage{colortbl}
\usepackage{amsmath}
\usepackage{amsfonts}
\usepackage{algpseudocode}
\usepackage[normalem]{ulem} %sout
\usepackage{xspace}
\usepackage{multirow} % Fancy tables
\usepackage{balance} % Reference balancing
\usepackage{fancyvrb} % Verbose mode in footnotes
\usepackage[export]{adjustbox} % Aligning differently-sized subfigures
\usepackage{subcaption} % Subfigures etc.
\usepackage{caption}
\usepackage{url}
\usepackage{xcolor}
\usepackage{tikz} % Import the tikz package
\usetikzlibrary{automata} % Import library for drawing automata
\usetikzlibrary{positioning} % ...positioning nodes
\usetikzlibrary{arrows} % ...customizing arrows
\usetikzlibrary{calc} % ...fancy edge calculations
\usetikzlibrary{patterns} % ...shading
\usetikzlibrary{snakes} % ...snakes
 
\usepackage[group-separator={,}]{siunitx}
\VerbatimFootnotes % Allow \verb in footnotes

\usepackage{cleveref}
\crefformat{section}{\S#2#1#3}
\crefname{figure}{Figure}{Figures}
\crefname{table}{Table}{Tables}
\crefname{listing}{Listing}{Listings}
\crefname{theorem}{Theorem}{Theorems}
\crefname{thm}{Theorem}{Theorems}
\crefname{lemma}{Lemma}{Lemmata}
\crefname{equation}{Eqt.}{Eqts.}
\crefformat{Grammar}{Grammar #1}

\usepackage{listings}
 
\usepackage{newfloat}
\usepackage{framed}
 
\DeclareFloatingEnvironment[
  fileext   = logr,% don't use log here!
  listname  = {List of Grammars},
  name      = Grammar,
  placement = htp
]{Grammar}

\ifSPACEHACK
\fi
\usepackage[ruled,vlined]{algorithm2e}
 
\usepackage{nicefrac}

\begin{document}

%%\title{Supporting the Conduct of Bug Studies: Reflections and Future Directions}

\title{Reflections on Software Failure Analysis}

%% Bug Studies are very relevant to achieving reliable software
%% Conducting Bug Studies is Difficult
%% Our results identifies various challenges and resulting problems
%% How can we better support researchers in conducting bug studies

\author{Paschal C. Amusuo}
\orcid{0000-0003-1001-525X}
\affiliation{%
  \institution{Purdue University}
  \country{USA}
}
\email{pamusuo@purdue.edu}

\author{Aishwarya Sharma}
\orcid{0000-0003-4033-3224}
\affiliation{%
  \institution{Purdue University}
  \country{USA}
}
\email{sharm234@purdue.edu}

\author{Siddharth R. Rao}
\orcid{0000-0001-9512-2593}
\affiliation{%
  \institution{Purdue University}
  \country{USA}
}
\email{rao147@purdue.edu}

\author{Abbey Vincent}
\orcid{0000-0003-1922-0276}
\affiliation{%
  \institution{Purdue University}
  \country{USA}
}
\email{vincen17@purdue.edu}

\author{James C. Davis}
\orcid{0000-0003-2495-686X}
\affiliation{%
  \institution{Purdue University}
  \country{USA}
}
\email{davisjam@purdue.edu}

%\renewcommand{\shortauthors}{Paschal Amusuo, Aishwarma Sharma, Siddhart Rao, Abbey Vincent, James Davis}

\newcommand\PA[1]{\textcolor{red}{#1}}

\begin{abstract}

  Failure studies are important in revealing the root causes, behaviors, and life cycle of defects in software systems. These studies either focus on understanding the characteristics of defects in specific classes of systems or the characteristics of a specific type of defect in the systems it manifests in. Failure studies have influenced various software engineering research directions, especially in the area of software evolution, defect detection, and program repair.
  
  In this paper, we reflect on the conduct of failure studies in software engineering. We reviewed a sample of 52 failure study papers. We identified several recurring problems in these studies, some of which hinder the ability of the engineering community to trust or replicate the results. Based on our findings, we suggest future research directions, including identifying and analyzing failure causal chains, standardizing the conduct of failure studies, and tool support for faster defect analysis.
    
\end{abstract}


%\ccsdesc[300]{Software and its engineering~Software post-development issues}


\maketitle

\section{Introduction}

The study of failures is integral to the success of engineered systems~\cite{petroski_design_1994}. In software engineering, failure studies describe the characteristics of defects in software systems. These studies, otherwise known as bug studies, are either tailored toward understanding the characteristics of defects in specific classes of systems (\eg web systems~\cite{chen_understanding_2019}, Android apps~\cite{linares-vasquez_empirical_2017}, or embedded systems~\cite{makhshari_iot_2021}) or the characteristics of specific classes of defects (\eg performance \cite{li_understanding_2021}, concurrency \cite{fonseca_study_2010}, or security \cite{mazuera-rozo_android_2019}). These studies are designed to reveal the root causes of these defects, their manifestation, impact, fix characteristics, and life-cycle.

\begin{figure}
 \includegraphics[width=0.90\columnwidth]{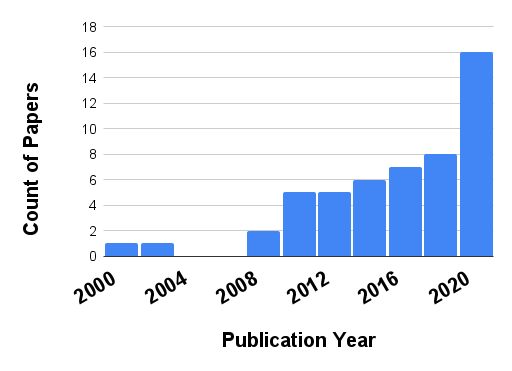}
 \captionof{figure}{
    The distribution of failure studies by year published.
 }
 \label{fig:published1}
 %\vspace{-0.5cm}
\end{figure}

Over the last decade, the number of failure studies has steadily increased (\Cref{fig:published1}). These studies have influenced research into software testing \cite{humbatova_deepcrime_2021}, defect detection \cite{dilley_bounded_2020}, and repair techniques ~\cite{ocariza_jr_vejovis_2014}.

\begin{figure*}[h]
 \centering
 \includegraphics[width=0.90\linewidth]{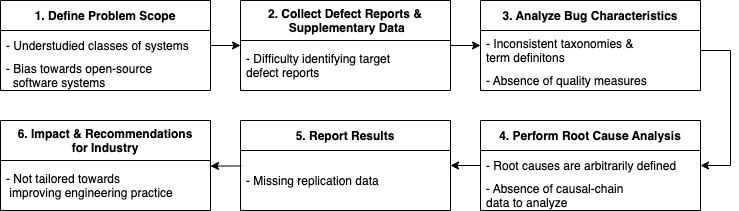}
 \captionof{figure}{
    Idealized model of software engineering failure study that our study identified flaws in.
 }
 \label{fig:failure-study-model}
 %\vspace{-0.5cm}
\end{figure*}

In this paper, we reflect on the conduct of software failure analysis research over the last 20 years. Using a systematic literature review, we identified several flaws and challenges that affect this research direction. Following the flaws and challenges we identified, we discussed future research directions that the software engineering community can embark on, to aid the conduct of these failure studies. Our research directions are focused on attempting to answer various questions relevant to the efficient conduct and impact of failure studies.

%% Cite papers that complain of the difficulty of conducting bug studies
%% 

%% Talk about bugs, bug reports and failure studies

\section{Idealized Failure Study Model}

Failure studies are research focused on understanding the characteristics and causes of failures in engineered systems~\cite{liblit_building_2002} \cite{ubani_study_2013}. In software engineering, these studies commonly consider defects. 

This section presents an idealized model of the failure study process in software engineering. We derived this model by reviewing steps currently taken to conduct software failure studies, complemented with failure studies conducted in other engineering disciplines \cite{fox_2001}. We used this model to analyze and review various failure studies reported in the software engineering literature.

\Cref{fig:failure-study-model} shows the various stages of this idealized model, which is applied across engineering disciplines. First, the project scope is defined. This usually involves identifying what class of defects to study, the system to study, and how the target defects and system would be identified. Then the defect reports and other relevant data are collected and reviewed. The investigators use the information extracted to analyze the characteristics of the various defects, such as how they manifest, their impact, their life cycle, etc. In addition to this, the investigators can also perform a root cause analysis to determine the probable root cause and contributing causes of the defects. Once the study is completed, investigators report their results and discuss their implications. This report should also contain their analyzed data to aid replicability by other investigators. To ensure that practitioners learn from the results of the study, it behooves the investigators to provide recommendations to these practitioners while also working with them to validate the impact of their results and recommendations.

The figure also depicts common shortcomings of the existing studies in software engineering literature at various stages. We discuss these shortcomings in the next section.

\section{Flaws in Failure Study Methods}

This section presents the flaws we identified in this research direction, as practiced in software engineering. 

\subsection{Methodology}
\label{section:methodology}

We first searched the proceedings of prominent software engineering conferences (ICSE, ESEC/FSE, ASE) and journals (IEEE TSE, ESEM, JSS) and manually identified failure study papers. The results helped us define our search phrase.\footnote{Our final search query was "(empirical OR comprehensive OR taxonomy OR characteristics) AND (bug OR bugs OR faults OR defects OR failures OR vulnerabilities) AND (study OR review)"} We used this phrase to search scholarly databases (Google Scholar, IEEE Xplore, ACM Digital Library). This search yielded 92 candidate papers. Working in teams of 2, we manually reviewed the abstract of these papers, identified and selected 52 papers that studied and characterized defects in software, and were published in peer review venues. 

We reviewed the selected papers and collected data related to the various stages outlined in \cref{fig:failure-study-model}. We analyzed the data extracted and identified the flaws discussed in the next subsection.

To ensure the quality of our results, we had multiple authors independently perform data extraction on a sample of 20 papers. We computed the Cohen kappa score on this sample as 0.763, which shows substantial agreement \cite{landis_measurement_1977}. Subsequently, the authors continued the data extraction independently while one more experienced author reviewed the data extracted by the other authors. 

\textbf{Threat to validity}: We sampled only 52 failure studies, which may not have included all relevant failure studies. But we believe this sample is representative, and our findings are valid and relevant. The sample was selected through a methodological process, as discussed above. We also included recent papers published in prominent venues to ensure our findings were relevant to the current peer-reviewed conduct. Also, each of the flaws we identified was prevalent in over half of the sample of papers studied. Finally, while some of the flaws identified may seem obvious, we are the first to present empirical evidence of their existence while suggesting research directions to manage them.

\subsection{Recurring Flaws}

\subsubsection{Bias towards Open-source Software: }
\label{section:bias-open-source}

Investigators conducting failure studies are biased toward studying defects in open-source software (first row of \cref{tab:analysis}). This is usually because open-source software has publicly available code, documentation, and complete evolution history. Unfortunately, focusing on only open-source software may be inconsistent with the investigator's goal, ultimately aiding software engineering practice beyond open-source.

Prior research has investigated and reported differences between open-source and commercial software \cite{mockus_case_2000} \cite{paulson_empirical_2004} \cite{boulanger_open-source_2005}. Mockus~\etal~\cite{mockus_case_2000} showed that the post-release defect density for Apache was significantly different compared to 4 commercial projects. Paulson~\etal~\cite{paulson_empirical_2004} reported that more defects are being found and fixed in open-source software, which may have contributed to the high defect density reported in \cite{mockus_case_2000}. Boulanger \cite{boulanger_open-source_2005} identified differences between the software development practices for open-source and commercial software projects. In open-source software, defects are usually reported by customers, unlike in commercial software. This could also affect the kinds of defects analyzed by failure studies.  As a result, the results from these failure studies that studied open-source software may not generalize to commercial environments.

\begin{table}[t]
\caption{
    Table showing further failure study analysis.
    }
\label{tab:analysis}
\begin{center}
\begin{small}
\begin{sc}
\begin{tabular}{lcccr}
\toprule
Analysis & Yes & No\\
\midrule
Papers that studied defects in proprietary software & 3 & 49 \\
Papers that reused taxonomies from literature   & 10 & 42\\
Papers that reported the use of any tool & 12 & 40\\
Papers that made practitioner-relevant contributions & 14 & 38 \\
\bottomrule
\end{tabular}
\end{sc}
\end{small}
\end{center}
\end{table}

\subsubsection{Root Causes are Subjectively Identified: }
\label{section:root-causes}

%% Change the "very few" to actual percentage value

%% Define what root cause should be
%% Root cause provide the why
%% Root causes should reveal systemic causes

\begin{figure}
 \includegraphics[width=0.90\columnwidth]{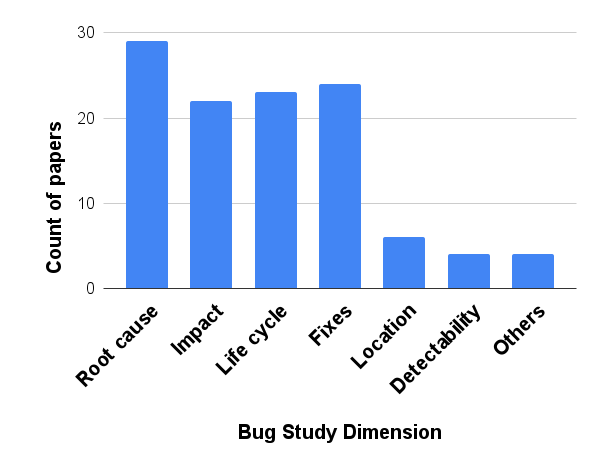}
 \captionof{figure}{
    Research questions investigated by failure studies.
 }
 \label{fig:bug-dimensions}
 %\vspace{-0.5cm}
\end{figure}

Root cause analysis is the most common aspect of defects considered by failure studies (\cref{fig:bug-dimensions}). However, only one paper~\cite{makhshari_iot_2021} reported using a root cause analysis methodology to identify these root causes. According to Paradies \etal~\cite{paradies_root_1988}, root causes should be basic causes that are within the ambit of management to fix. Gangidi \cite{gangidi_systematic_2018} also explained that a systematic root cause analysis methodology should reveal deeper systemic causes (\eg policies, practices, management decisions). 

The root causes identified by the failure studies we reviewed mostly represent technical flaws and do not correspond with any of these definitions. Wang~\cite{wang_exploratory_2021} identified root causes such as misuse of mathematical formulas, inconsistency between hardware and software, and improper handling of parameters. While these are the immediate causes of the reported defects, they are neither `basic' nor systemic. Deeper investigations into defects caused by hardware/software inconsistency may reveal underlying causes such as poor documentation, which may also have been attributed to the absence of documentation guidelines. As another example, Gunawi \etal~\cite{gunawi_what_2014} identified \textit{data races} as one of the root causes of data inconsistency in cloud systems, but deeper analysis might have also revealed other underlying factors that led to these data races. If papers conducted a deeper root cause analysis, their results could be more helpful to practitioners and engineering teams.

\subsubsection{Inconsistent Defect Taxonomies: }
\label{section:inconsistent-taxonomies}

Failure studies attempt to characterize the defects in software systems to aid their analysis. Our results, as shown in the second row of \cref{tab:analysis}, show that most failure studies invent the taxonomies they use for this characterization, even when they study the same class of defects. For example, Cao \etal~\cite{cao_characterizing_2021} characterized performance bugs in deep learning systems using a self-generated taxonomy but could have adapted taxonomies from prior research on performance bugs \cite{liu_characterizing_2014} \cite{mazuera-rozo_investigating_2020} \cite{zaman_qualitative_2012}. As a result, it becomes difficult to compare the distribution of performance defects in \cite{cao_characterizing_2021} and earlier works such as \cite{mazuera-rozo_investigating_2020}.

We also found disagreement in the interpretation of terms in the taxonomy when investigators choose to reuse taxonomies from earlier studies. For example, Tan \etal~\cite{tan_bug_2014} reported they reused the taxonomy defined by Sullivan \etal~\cite{sullivan_comparison_1992} but acknowledged that the definition of \textit{semantic bugs} between the two studies may be different, accounting for the huge discrepancy between the percentage of semantic bugs found by the two papers.

\subsubsection{Non-integration of Practicing Software Engineers in the Study: }
\label{section:non-integration-engineers}
Our review of failure study papers shows that practitioners are not included during the conduct of these studies. Investigating the perspectives of the software engineers who create or fix these defects can be helpful in providing insights into the causes and characteristics of these defects.

Furthermore, failure study papers are focused on enabling software engineering research but fail to make contributions that are relevant to software engineers. According to the fourth row of \cref{tab:analysis}, only 27\% of reviewed papers proposed recommendations pertinent to current software engineering practices. Mantyla \cite{mantyla_what_2009} provided guidelines for conducting code and documentation reviews. Sun \cite{sun_toward_2016} made recommendations for generating test cases for compilers. Others only discussed the research implications of their work. This is contrary to failure studies in other disciplines whose results recommended changes in practitioners' practices \cite{fox_2001} \cite{petroski_design_1994} \cite{reason_human_1990} \cite{reason_1997_organizational}. With an increased focus on improving engineering practice, the results and recommendations from these studies could reduce the occurrence of defects, which would significantly increase software engineers' productivity..

\subsubsection{Defects in Embedded/IoT Systems are Understudied: } 
\label{section:understudied-defects}

\begin{figure}
 \includegraphics[width=0.90\columnwidth]{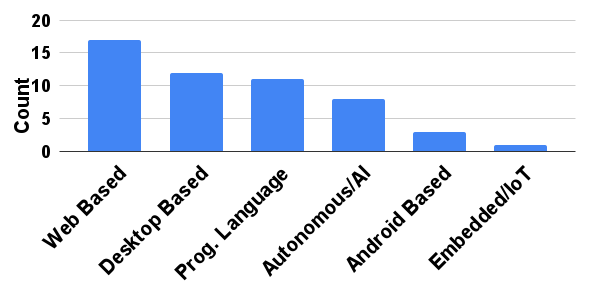}
 \captionof{figure}{
    Distribution of failure studies by system type.
 }
 \label{fig:systems-studied}
 %\vspace{-0.5cm}
\end{figure}

From our results, we observed that the software engineering community is biased towards failure studies on web-based and desktop-based systems, while embedded/IoT systems are still understudied. As shown in \cref{fig:systems-studied}, embedded/IoT systems accounted for only two papers, while web-based systems (\eg browsers) had 16 and desktop-based systems (\eg compilers) had 12. Embedded systems power our airplanes, vehicles, and industries and deserve additional attention. 

\subsubsection{Miscellaneous Flaws: }
\label{section:miscellaneous}

In addition to the primary flaws discussed above, we summarize three more issues.

Inconsistent quality measures: Defect analysis is subjective, and single-author investigation methods are untrustworthy. Of the 52 papers reviewed, only 19 studies had multiple authors independently analyze the data. Hence, the results of most studies are untrustworthy without the use of quality control measures.

Absence of replicability data: Only 11 papers included links to their replication package; 3 of these were inaccessible.

Missing tool support: Failure studies are time-consuming and lack tool support. Leesatapornwongsa \etal~\cite{leesatapornwongsa_taxdc_2016} and Shen \etal~\citep{shen_comprehensive_2021} reported that it took them 15 and 24 months to conduct their study. Yet, according to the third row of \cref{tab:analysis}, only 23\% of failure studies reported using any tool in their study. These studies require investigators to analyze and categorize hundreds of defect reports manually. When studying a specific class of defects, these investigators rely on only keyword matching to filter prospective defect reports and need to go through each filtered report to identify and remove false positives. Mazuera-rozo \etal~\cite{mazuera-rozo_investigating_2020} identified 1,010 commits using keyword matching, and after manual analysis by two authors, only 20\% (204 commits) were true positives.

%% Also, talk about an increased focus on open-source rather than commercial software. Include recommendations to develop frameworks enabling industry practitioners to conduct these failure analyses.
%% Use bolded inline topics to aid readability

\section{A Research Agenda}

%Motivated by our results, we discuss four research directions.

\subsection{Defect Causal Chains}

To effectively identify the root causes of defects, as discussed in \cref{section:root-causes}, we suggest investigators use additional sources that provide more information about the causal chain of the defect. It is uncertain if analysis of pull request comments, meeting logs, design documents, or other artifacts will be helpful. Still, these documents can provide more insights into the reason behind the codes written by the developers. The research community can conduct further research to determine which artifacts would be more helpful and how investigators can adequately analyze them to identify the root causes of defects.

In addition, software engineers have no standard approach to documenting design or implementation decisions or efforts. While standards such as ISO/IEC/IEE 12207 require detailed documentation by the software engineers, Agile methodologies \cite{agile_manifesto} \cite{beck_2005} recommend less comprehensive documentation. Hence, this presents another challenge as there is no guarantee that these documents will be available for analysis. The research results can also inform engineering teams what documentation needs to be maintained if they want to learn from their failures.

\subsection{Standardizing the Conduct of Failure Studies}

As we discussed in \cref{section:inconsistent-taxonomies}, there are inconsistencies in the conduct of failure studies. We suggest two ways to standardize the conduct of these studies.
First, add a standard for failure analysis to the SIGSOFT empirical standards \cite{ralph_full_2020} to note the quality measures, replication packages, and expected general guidelines for conducting a failure study. 
Second, we suggest the development of a defect-type taxonomy map for software defects, similar to the Common Weakness Enumeration (CWE) used for categorizing security vulnerabilities. Such a map would contain a taxonomy of common defect types. It can be extensible that investigators conducting failure studies for a specific system or defect classes can build upon existing taxonomies with defect type categories particular to the class of system being investigated rather than inventing a new taxonomy. This map would ensure that the results of all failure studies are comparable, which will improve the generalizability of research influenced by the results.

\subsection{Increased Impact on Engineering Practices}

Following the bias reported in \cref{section:bias-open-source}, we propose increased research emphasis on replicability studies to verify if failure studies conducted on open-source software also hold for commercial software. We also suggest increased collaboration between investigators of failure studies and software engineering companies, which would provide these investigators access to defect reports of commercial software. This collaboration would ensure that failure studies' results influence research, which would also be relevant to practitioners in these companies.

We also recommend that, in addition to providing research directions, software failure studies provide recommendations to engineering teams that will reduce the occurrence of defects and the time to debug and fix reported defects. This is akin to failure analysis in other engineering disciplines, such as in the NTSB, where such studies have led to various changes in engineering, management, and regulatory practices \cite{fox_2001}.

\subsection{Tool Support for Faster Defect Analysis}
\label{section:tool-support}

With the challenge of missing tool support discussed in \cref{section:miscellaneous}, we recommend the research and development of tools that would aid the conduct of these studies. Natural Language Processing (NLP) techniques have become increasingly helpful in understanding the semantic meaning of documents, summarizing, and extracting useful information from documents. They have successfully been used to identify defects in requirement documents \cite{rosadini_using_2017}, identify duplicate defect reports \cite{runeson_detection_2007}, extract tasks and user stories from app store reviews \cite{guo_caspar_2020}, and summarize defect reports \cite{rastkar_automatic_2014} \cite{rastkar_summarizing_2010}. Hence, the research community can easily explore the use of NLP to identify target defect reports, characterize the defects in them and extract other relevant information about the defect (\eg consequence, manifestation behavior, component affected) from these reports. While using NLP can not replace the need for expertise-based human analysis, automating the above-listed tasks would significantly reduce the time the investigators spend conducting manual analysis.

\section{Conclusion}
In this paper, we reflect on the conduct of failure studies in software engineering by surveying 52 published failure study papers. We identified eight recurring flaws that have marred the conduct of failure studies. These flaws impede the correctness, reliability, and impact of the reported results of these studies.

Motivated by these challenges, we identify various ways the research community can support the conduct of these failure studies. We encourage further research on identifying and analyzing causal chains for defects and tool support to simplify defect analysis while recommending efforts to standardize the conduct of failure studies. With these steps, software failure studies may improve software engineering quality.

\section*{Data Availability}

Our artifact can be found at \url{https://doi.org/10.5281/zenodo.7041931}. This spreadsheet contains our analysis of the failure study papers we surveyed.

%\raggedbottom
%\balance

\bibliographystyle{ACM-Reference-Format}
\bibliography{bib/sample-base, bib/others}

\end{document}